# Electronic and elastic properties of the new 7.5K superconductor $Nb_2InC$ from first principles

I.R. Shein, A.L. Ivanovskii[*]

*Institute of Solid State Chemistry, Ural Branch of the Russian Academy of Sciences, 620990, Ekaterinburg, Russia*

First-principles calculations were performed to investigate electronic and elastic properties of the newly discovered 7.5K superconductor: layered $Nb_2InC$. As a result, electronic bands, total and site-projected *l*- decomposed DOS at the Fermi level, shape of the Fermi surface for $Nb_2InC$ were obtained for the first time. Besides, independent elastic constants, bulk modulus, compressibility, shear modulus, Young's modulus, Poisson's ratio together with the elastic anisotropy parameters and indicator of brittle/ductile behavior of $Nb_2InC$ were evaluated and analyzed in comparison with the available data.

*PACS:* 71.15.Mb; 74.25.Jb; 74.25.Ld

*Keywords:* Superconducting $Nb_2InC$; *ab initio* calculations; Electronic properties; Elastic properties

---

[*] Corresponding author.
*E-mail address:* ivanovskii@ihim.uran.ru (A.L. Ivanovskii).

The recent discovery of high-temperature superconductivity (reviews [1-4]) in so-called FeAs systems has stimulated much activity in the search for new related layered superconducting materials.

To these layered materials belong so-called nanolaminates, termed often also as $M_{n+1}AX$ (MAX) phases, where $M$ is a transition metal, A is a $p$ element (Al, Si, Ge, Ga etc) and X is C or N. These phases exhibit a unique combination of physical properties, which are characteristic both of metals and ceramics. Like metals, $M_{n+1}AX_n$ phases are electrically and thermally conductive, plastic, and damage tolerant; similarly to ceramics, they are lightweight, elastically rigid, and maintain strength to high temperatures etc., see review [5].

The metallic-like nature of $M_{n+1}AX_n$ phases allows us to consider these materials as potential superconductors. Indeed, among ~ 60 synthesized $M_{n+1}AX_n$ phases [5], until recently bulk superconductivity was discovered for five systems: $Mo_2GaC$ ($T_C$ ~ 4K [6]), $Nb_2SC$ ($T_C$ ~ 5K [7]), $Nb_2SnC$ ($T_C$ ~ 7.8K [8]), $Nb_2AsC$ ($T_C$ ~ 2K [9]) and $Ti_2InC$ ($T_C$ ~ 3K [10]). Very recently a superconducting transition at $T_C$ ~ 7.5K was reported for the sixth MAX phase - $Nb_2InC$ [11].

Unlike some other superconducting MAX phases, for which their electronic properties have been examined ($Nb_2AsC$ [9,12], $Nb_2SC$ [12-14], $Nb_2SnC$ [14,15], $Ti_2InC$ [16-18]), no data about the electronic spectrum of $Nb_2InC$ are available at present.

In this Letter, we report the results of the first-principles calculations of the electronic band structure, densities of states (DOSs) and the Fermi surface for the newly discovered superconductor $Nb_2InC$. Furthermore, the elastic properties are of great interest for the material science of superconductors (SC's); for example the elastic constants can be linked to such important physical parameters of SC's as the Debye temperature $\Theta_D$ and the electron-phonon coupling constant λ. Besides, mechanical properties appear important for technology and advanced applications of superconducting materials. Therefore we have evaluated also for examined SC $Nb_2InC$ the most important elastic parameters, namely elastic constants, bulk modulus, compressibility, shear modulus, Young's modulus and Poisson's ratio; in addition the elastic anisotropy parameters and indicator of brittle/ductile behavior of this crystal have been estimated.

The considered Nb$_2$InC adopts the hexagonal structure with space group *P*6$_3$/*mmc* (No. 194), where the blocks of Nb carbide [NbC] (formed by edge-shared Nb$_6$C octahedra) are sandwiched between In atomic sheets. The Wyckoff positions of atoms are – carbon: 2*a* (0, 0, 0), In: 2*d* (1/3, 2/3, 3/4), and Nb atoms: 4*f* (1/3, 2/3, $z_{Nb}$). The structure is defined by lattice parameters *a* and *c*, and the internal parameter $z_{Nb}$, see [5].

The electronic band structure of Nb$_2$InC was examined by means of the full-potential method with mixed basis APW+lo (LAPW) implemented in the WIEN2k suite of programs [19]. The generalized gradient correction (GGA) to exchange-correlation potential of Perdew, Burke and Ernzerhof [20] was used. The basis set inside each MT sphere was split into core and valence subsets. The core states were treated within the spherical part of the potential only, and were assumed to have a spherically symmetric charge density confined within MT spheres. The valence part was treated with the potential expanded into spherical harmonics to *l* = 4. The valence wave functions inside the spheres were expanded to *l* = 10. The plane-wave expansion with $R_{MT} \times K_{MAX}$ was equal to 7.0, and *k* sampling with 14×14×4 *k*-points mesh in the full Brillouin zone was used. The MT sphere radii were chosen to be 1.6 a.u. for carbon, 2.2 a.u. for In, and 2.5 a.u. for Nb. In addition, for the calculations of the elastic parameters of Nb$_2$InC the we have employed the Vienna *ab initio* simulation package (VASP) in projector augmented waves (PAW) formalism [21,22]. Exchange and correlation were described by a nonlocal correction for LDA in the form of GGA [20]. The kinetic energy cutoff of 500 eV and k-mesh of 16×16×6 were used. The geometry optimization was performed with the force cutoff of 2 meV/Å.

These two DFT-based codes are complementary and allow us to perform a complete investigation of the declared properties of Nb$_2$InC.

As the first step, the equilibrium lattice constants (*a* and *c*) for Nb$_2$InC were calculated with full structural optimization including internal $z_M$ parameters. The results obtained are presented in Table and appear to be in reasonable agreement with the available theoretical and experimental data [11,23-25].

The band structure for Nb$_2$InC calculated at equilibrium lattice parameters, as well as the total and site-projected *l*- decomposed densities of states (DOSs) are shown in Fig. 1 and 2, respectively. It is found that the lowest valence bands in the range from -12.5 eV to -11.0 eV below Fermi level (E$_F$ = 0 eV) arise mainly from quasi-core C 2*s* states and are separated by a wide forbidden gap (at about 2 eV) from the higher valence bands, which are located in the energy range from -8.9 eV to $E_F$. In turn, these bands can be divided into two groups. The lowest of them, located in the range from -8.9 eV to -3.6 eV, originate mainly from mixed C $2p$ and Nb $4d$ states with admixture of In $5s$ states, and are also separated by a indirect gap from the highest occupied near-Fermi bands, see Fig. 1. The DOS includes an intense peak located around -4 eV. This peak contains strongly hybridized Nb $4d$ - C $2p$ states, which are responsible for the covalent Nb-C bonding inside [NbC] blocks. In addition, the overlap of Nb-In valence states is visible, *i.e* covalent interaction occurs between [NbC] blocks and In sheets. The near-Fermi bands are mainly of Nb $4d$ type with small admixtures of carbon and indium states.

The overall bonding picture in $Nb_2IC$ may be described as a mixture of covalent, metallic, and ionic interactions. In addition to the above mentioned covalent Nb-C and Nb-In bonds, the metallic-like Nb-Nb bonding occurs owing to overlapping of the near-Fermi Nb $4d$ states, and the ionic contribution is due to the difference in electronegativity between the comprising elements: Nb (1.60), C (2.55) and In (1.78).

In the vicinity of the Fermi level, Nb $4d$ states dominate and should contribute to the conduction properties of $Nb_2InC$. According to our calculations, the total DOS at the Fermi level $N^{tot}(E_F)$ = 1.81 states/eV·f.u., where the main contribution is from Nb $4d$ electrons: $N^{Mo4d}(E_F)$ = 1.07 states/eV·f.u., *i.e.* ~ 59 %. Carbon makes almost no contribution to the DOS at the Fermi level ($N^{C2p}(E_F)$ = 0.05 states/eV·f.u., *i.e.* < 3 %), and therefore does not participate in conductivity. The contributions from the valence states of In sheets ($N^{In\ s,p,d}(E_F)$ = 0.06 states/f.u.) are small enough too - about 3%.

The Fermi surface of $Nb_2InC$ has enough complex topology and contains electronic- and hole-like sheets, which are centered along the Γ-A direction as well as two-dimensional-type sheets parallel to the $k_z$ direction, centered at the lateral sides of the Brillouin zone along the L-M direction, see Fig. 3. The Fermi surface of $Nb_2InC$ is formed mainly by the low-dispersive bands from [NbC] layers, which should be responsible for superconductivity for $Nb_2InC$.

Next, let us discuss the elastic parameters for $Nb_2InC$. The values of five independent elastic constants for hexagonal crystals ($C_{11}$, $C_{12}$, $C_{13}$, $C_{33}$, and $C_{44}$) as well as the bulk *B* and shear *G* moduli (as obtained in Voigt-Reuss-Hill scheme [26]) are given in Table. Further, the calculated moduli *B* and *G* allow us to obtain the Young's modulus $Y$ and Poisson's ratio $v$ as: $Y = 9BG/(3B + G)$ and $v = (3B - 2G)/\{2(3B + G)\}$. The above elastic parameters presented in Table allow us to make the following conclusions:

(i) For $Nb_2InC$, all of $C_{ij}$ constants are positive and satisfy the generalized criteria [27] for mechanically stable crystals: $C_{44} > 0$, $C_{11} > |C_{12}|$, and $(C_{11} + C_{12}) C_{33} > 2C_{13}^2$.

(ii). For $Nb_2InC$ $B > G$; this implies that the parameter limiting the mechanical stability of this material is the shear modulus.

(iii). According to Pugh's criteria [28], a material should behave in a ductile manner if $G/B < 0.5$, otherwise it should be brittle. In our case $G/B = 0.436$, i.e. according to this indicator $Nb_2InC$ will behave as a ductile material. An additional argument for the ductile behavior of this superconductor follows from the calculated Poisson's ratio $v$. Indeed, these values for brittle materials are small ($v \sim 0.1$), whereas for ductile metallic materials $v$ is typically 0.33 [29]. In our case, $v \sim 0.31$ is close to this limit.

(iv). The elastic anisotropy of crystals is an important parameter for material science of superconductors since it correlates with the possibility of appearance of microcracks in materials, see [30]. There are different ways to represent the elastic anisotropy of crystals, for example, by using the calculated $C_{ij}$ constants. For this purpose, so-called shear anisotropy ratio $A = 2C_{44}/(C_{11} - C_{12})$ is often used. The shear anisotropic factors may be obtained as a measure of the degree of anisotropy in the bonding between atoms in different planes [31,32]. In this approach, the shear anisotropic factor $A$ for the {100} shear planes between ‹011› and ‹010› directions is defined as: $A_1 = 4C_{44}/(C_{11} + C_{33} - 2C_{13})$. The magnitude of the deviation from $A, A_1 = 1$ is a measure of elastic anisotropy. From our data (Table) we can conclude that $Nb_2InC$ adopts the elastic anisotropy. Additionally, so-called percent shear anisotropy for polycrystalline materials [31] may be estimated as: $A_G = (G_V - G_R)/(G_V + G_R)$, where $A_G = 0$ correspond to elastic isotropy whereas the values of 100% correspond to the largest possible anisotropy. The values $G_V$ and $G_R$ are shear moduli as obtained by means of Voigt (V) [33] and Reuss (R) [34] schemes. In our case, $A_G = 0.0233$. Finally, the parameter $k_c/k_a = (C_{11} + C_{12} - 2C_{13})/(C_{33} - C_{13})$ is used, which expresses the ratio between linear compressibility coefficients of hexagonal crystals [35]. The data obtained $k_c/k_a = 0.780$ demonstrate that the compressibility for $Nb_2InC$ along the $c$ axis is smaller than along the $a$ axis.

In summary, the first-principles FLAPW-GGA and VASP methods have been used for study of the electronic and elastic properties of the newly discovered 7.5K superconductor $Nb_2InC$. We found that that the main contributions to the density of states at the Fermi level come from the Nb $4d$ states. The FS is formed mainly by the low-dispersive Nb $4d$ like bands from [NbC] layers, which should be responsible for superconductivity for $Nb_2InC$.

The evaluated elastic parameters allow us to conclude that superconducting $Nb_2InC$ is mechanically stable crystal; the parameter limiting the mechanical stability of this material is the shear modulus. In addition, $Nb_2InC$ can be characterized as ductile material which will show elastic anisotropy. Finally, let us note that elastic parameters of the newly discovered superconductor $Nb_2InC$ are higher than for example for layered FeAs SCs, which are relatively soft materials ($B < 100$ GPa) [36,37], but are comparable with the same for some others layered SC's (such as YBCO, $MgB_2$, borocarbides, carbide halides of the rare earth metals *etc*) for which their bulk moduli do not exceed $B \leq 200$ GPa, see [36].

**Acknowledgments**
Financial support from the RFBR (Grant 09-03-00946-a) is gratefully acknowledged.

**Table.** Calculated lattice parameters ($a$, $c$, in Å), ratio $c/a$, internal parameter ($z_{Nb}$), cell volume ($V_o$, in Å$^3$) elastic constants ($C_{ij}$, in GPa), bulk modulus ($B$, in GPa), compressibility ($\beta$, in GPa$^{-1}$), shear modulus ($G$, in GPa), the Young's modulus ($Y$, in GPa) Poisson's ratio ($v$) and elastic anisotropy parameters ($A$, $A_1$, $A_G$ and $k_c/k_a$, see text) for superconducting Nb$_2$InC as obtained from VASP calculations in comparison with others available data.

| $a$* | $c$ | $c/a$ | $z_{Nb}$ | $V_o$ |
|---|---|---|---|---|
| 3,1933 (3.17 [23]; 3.172 [11]; 3.137 [24]; 3.196 [25]) | 14,4952 (14.37 [23, exp]; 14.280 [24]; 14.47 [25]) | 4.5392 (4.647 [23]; 4.552 [24];) | 0.0821 (0.0830 [24]) | 128,01 |
| $C_{11}$ | $C_{12}$ | $C_{13}$ | $C_{33}$ | $C_{44}$ |
| 291,3 (363 [24]; 291 [25]) | 77,4 (103 [24]; 76 [25]) | 117,6 (131 [24]; 108 [25]) | 288,7 9 (306 [24]; 267 [25]) | 57,1 (148 [24]; 102 [25]) |
| $B$ | $\beta$ | $G$ | $G/B$ | $Y$ |
| 182,43 (195 [24]; 159 [25]) | 0,005482 | 79,63 (128 [24]; 99 [25]) | 0.436 | 208,54 (314 [24]; 247 [25]) |
| $v$ | $A$ | $A_1$ | $k_c/k_a$ | $A_G$ |
| 0,3095 (0.2319 [24]) | 0.534 | 0.662 | 0.780 | 0.0233 |

* in parentheses the available experimental [11,23] and theoretical [24,25] data are given. Calculations: [24] CASTEP (Cambridge Serial Total Energy Package) code; [25] – VASP code with GGA approximation using the PW91 functional.

# Figures

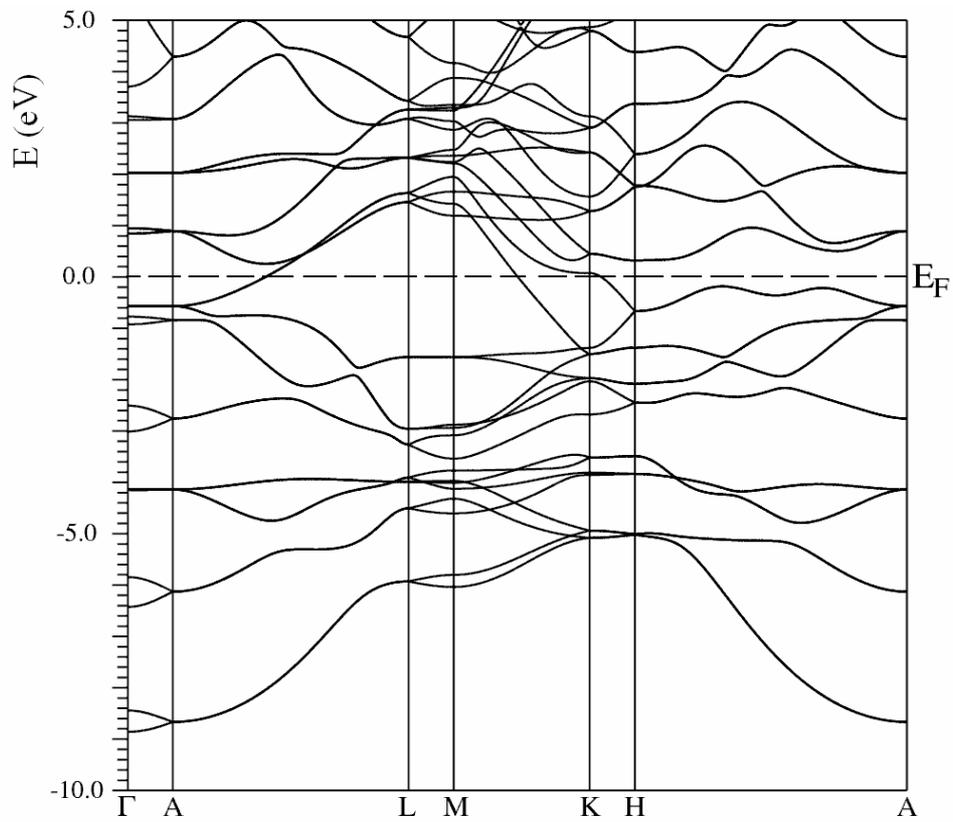

**Fig. 1.** Electronic bands of superconducting $Nb_2InC$ as obtained from FLAPW–GGA calculations.

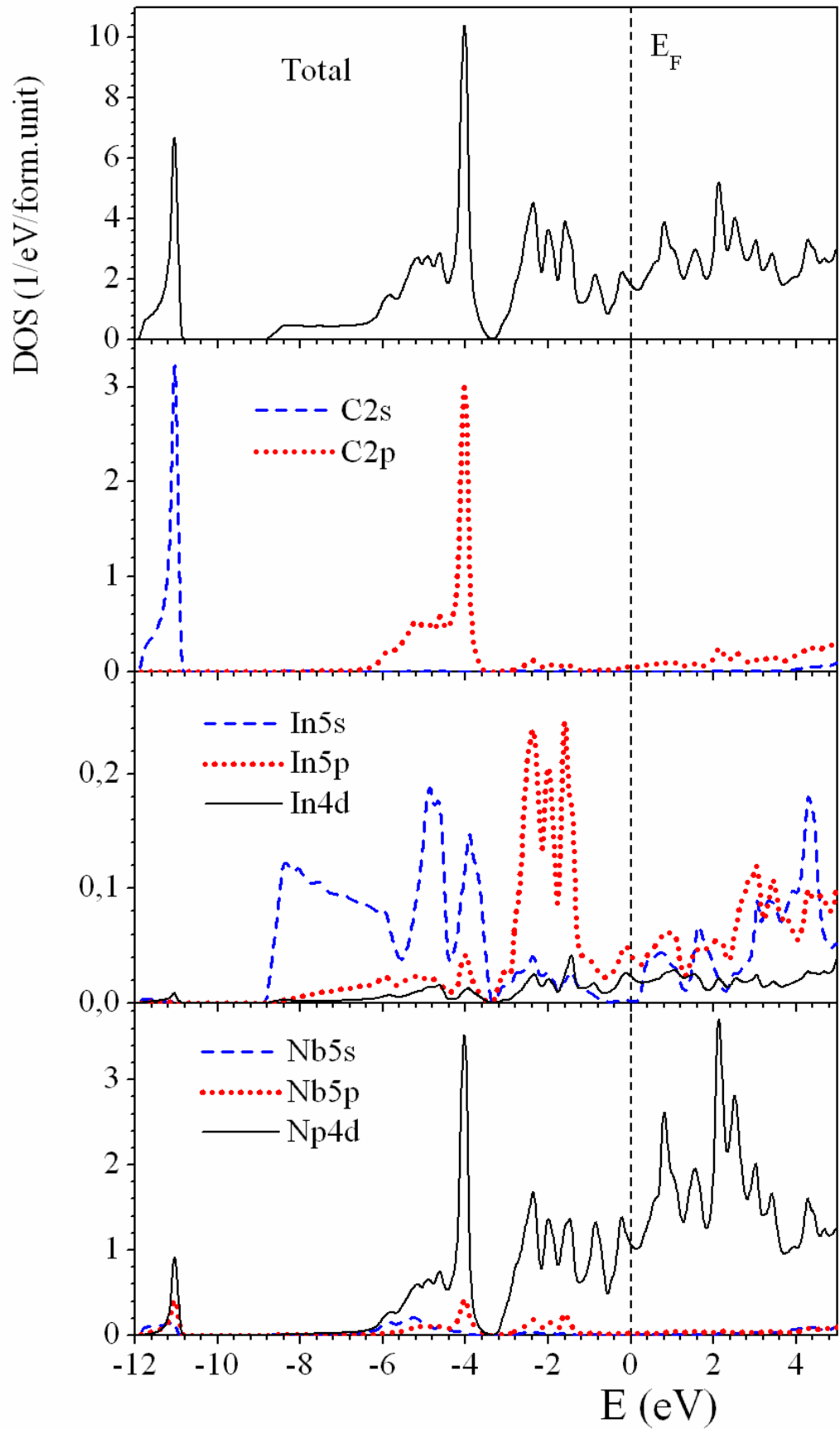

**Fig. 2**. Total and partial densities of states of Nb$_2$InC as obtained from FLAPW–GGA calculations.

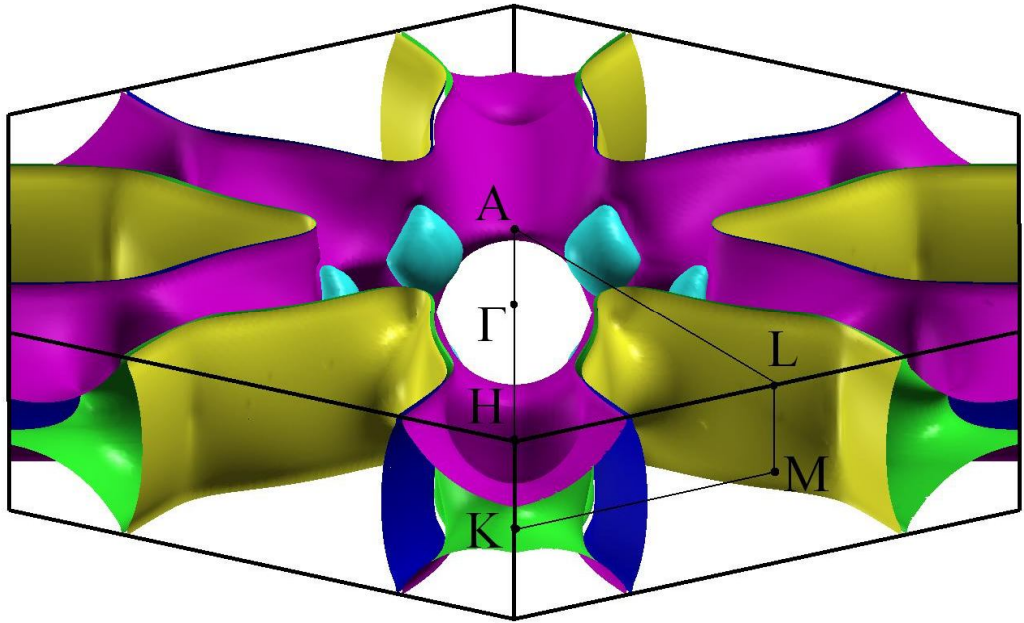

**Fig. 3.** The Fermi surface of $Nb_2InC$ as obtained from FLAPW–GGA calculations.